\documentclass[preprint]{aastex}

\shorttitle{Hard X-ray Excess in Radio-Quiet AGN}
\shortauthors{Tatum et al.}

\begin{document}

\title{The Global Implications of the Hard Excess II: Analysis of the Local population of Radio Quiet AGN}

\author{M.M. Tatum\altaffilmark{1,2}, T.J. Turner\altaffilmark{2}, L. Miller\altaffilmark{3}, J.N. Reeves\altaffilmark{2,4}, J. DiLiello\altaffilmark{2}, J. Gofford\altaffilmark{2,4}, A. Patrick\altaffilmark{4}, M. Clayton\altaffilmark{3}}

\altaffiltext{1}{NASA Goddard Space Flight Center, Code 662, Greenbelt, MD 20771, USA}

\altaffiltext{2}{Department of Physics, University of Maryland Baltimore County, Baltimore, MD 21250, USA}

\altaffiltext{3}{Dept. of Physics, University of Oxford, Denys Wilkinson Building, Keble Road, Oxford OX1 3RH, U.K.}

\altaffiltext{4}{Astrophysics Group, School of Physical and Geographical Sciences, Keele 
University, Keele, Staffordshire ST5 5BG, U.K}

\begin{abstract}
Active galactic nuclei (AGN) show evidence for reprocessing gas,
outflowing from the accreting black hole. The combined effects of
absorption and scattering from the circumnuclear material likely explains the `hard
excess' of X-ray emission above 20 keV, compared with extrapolation of
spectra from lower X-ray energies. In a recent {\it Suzaku} study, we
established the ubiquitous hard excess in hard X-ray-selected, radio-quiet type\,1 AGNs to
be consistent with reprocessing of the X-ray continuum an ensemble of
clouds, located tens to hundreds of gravitational radii from the
nuclear black hole. Here we add hard X-ray-selected, type\,2 AGN to extend our original
study and show that the gross X-ray spectral properties of the entire local
population of radio quiet AGN may be described by a simple unified
scheme.  We find a broad, continuous distribution of spectral hardness ratio
and Fe\,K$\alpha$ equivalent width across all AGN types,
which can be reproduced by varying the observer's sightline
through a single, simple model cloud ensemble, provided the radiative
transfer through the model cloud distribution includes not only 
photoelectric absorption but also 3D Compton scattering. Variation in
other parameters of the cloud distribution, such as column density or
ionisation, should be expected between AGN, but such variation is not
required to explain the gross X-ray spectral properties.

\end{abstract}


\section{Introduction}
X-ray production in Active Galactic Nuclei (AGN) is thought to arise from a corona of relativistic electrons situated above an accretion disk within tens of gravitational radii (r$_g$ = GM/c$^2$) of the black hole, enabling the observer to probe AGN on the smallest scales currently possible \citep[e.g.][]{rees77a}. These continuum photons would be emitted quasi-isotropically, depending on the shape and optical depth of the corona. Some of the continuum X-ray photons could pick up the imprint of gas along the line-of-sight or could illuminate and reflect from material out of the line-of-sight before reaching the observer. Either scenario results in broadband X-ray spectra comprising a combination of primary and reprocessed photons. In practice, separating the primary continuum and the reprocessed photons in X-ray data proves difficult with current observational data. In addition, the large uncertainties as to the form of the continuum, a result of not knowing the coronal conditions (see, e.g. \citealt{Fabian:1994gf}, and references therein), make the primary continuum difficult to isolate. 

There is a large amount of evidence showing that a complex absorber covers some of the sight-lines to local AGN. High resolution UV spectroscopy has revealed multi-layered, complex absorption to be a common phenomenon in AGN \citep[see][for a review]{Crenshaw:2003vn}. The detection of absorption features, such as  H- and He-like species of C, N, O, Ne, Mg, Al, Si, and S (e.g., \citealt{Kaspi:2002lr}), has shown that the signatures of complex absorption extend into the X-ray regime. The detection of deep Fe {\sc xxv} and Fe {\sc xxvi} absorption lines expanded the known range of X-ray absorbing column density in the local AGN population up into the Compton-thick regime \citep[e.g.,][]{Pounds:2003uq, Miller:2007fe,Turner:2008qy}. Further study of these absorption signatures in radio-quiet AGN revealed Fe {\sc xxv} and Fe {\sc xxvi} absorption lines to be common, in $\sim$40\% of the sources studied, with velocities up to 0.3\,c and column densities ranging from 21.5 $<$ log(N$_H$/ cm$^{-2}$) $<$  24.0 \citep[e.g.,][]{Tombesi:2010yq,Gofford:2013lr}. Supporting this picture of a Compton-thick, reprocessing wind, \citet{Tatum:2012fk} found, in a small sample of Seyfert 1 galaxies with little intrinsic absorption, that the moderately broad Fe K$\alpha$ emission line profile was consistent with production in a Compton-thick, accretion-disk wind of Solar abundances, arising tens to hundreds of r$_g$ from the black hole. 

Interestingly, changes in covering fraction of the X-ray absorber have been invoked to explain the marked spectral variability in some sources. For example, in a study of NGC 3516, \citet{Turner:2008qy} found that the emission and absorption features detected in {\it Chandra} HETG data confirmed the presence of four distinguishable zones of gas. The spectral variability in this source, observed on a timescale of $\sim$ 30 ks, was attributed to variations in the covering fraction of a layer of gas having column density {\it N}$_H$ $\sim$ 2 $\times$ 10$^{23}$ cm$^{-2}$ and ionization parameter log $\xi$ $\sim$ 2.2. 

Additional evidence for complex absorption has come from {\it Suzaku}. The {\it Suzaku} observations of NGC 4051, 1H 0419--577, PDS 456 and NGC 1365 revealed a marked excess of flux above 20 keV,  compared to that predicted from fits to data below 10 keV, dubbed a `hard excess'  
\citep[][respectively]{terashima09a,Turner:2009ys, Reeves:2009zt, Risaliti:2009lr}. In these sources, the high PIN-band flux (15--50 keV) was explained by the presence of a low ionization, Compton-thick absorber covering $>$ 70\% of the continuum source along the line-of-sight, extending the absorber complex to low ionization parameter, high column density zones. In order to obtain the true intrinsic X-ray luminosity of these sources, one must apply significant corrections for absorption and for Compton-scattering losses \citep[e.g.][]{Turner:2009ys, Reeves:2009zt}.  

\citet{Risaliti:2013lr} claimed to have ruled out absorption-dominated models in the {\it NuSTAR} observation of NGC 1365.  However, in the Compton-thick gas scenario, the amount of transmitted and scattered light from the Compton-thick, partial-covering gas has a significant effect in the X-ray band and is highly geometry-dependent.  \citet{Miller:2013fk} showed, using 3D radiative transfer calculations, that even simple considerations of the geometry of the partial-covering absorber can have a significant effect on X-ray spectra and that the simple slab models used by \citet{Risaliti:2013lr} did not form a valid  basis for ruling out  absorption models as a class. 
 
\citet[hereafter T13]{Tatum:2013lr} conducted an exploratory study of the strong excess of flux at energies $\ga$ 20 keV in the local type 1 AGN population, using the {\it Swift} Burst Alert Telescope (BAT) 58-month catalog\footnote{http://heasarc.gsfc.nasa.gov/docs/swift/results/bs58mon/} \citep{Baumgartner:2010fk}. The bandpass of BAT ($\sim$ 15-150 keV) allows a relatively unbiased survey of the X-ray sky for column densities up to $\sim$ 10$^{24}$ cm$^{-2}$. T13 selected all type 1 AGN, including intermediates up to type 1.9 and excluding known radio sources and LINERS, from the 58-month BAT catalog. This source list was cross-correlated with the sources in the {\it Suzaku} public archive, in order to obtain simultaneous medium (2--10 keV) and hard ($>$10 keV) X-ray data. Simultaneous data is required to avoid cross-calibration issues, as some AGN are known to vary above 10 keV \citep[e.g.][]{Miller:2008kx,Turner:2008qy}. Only sources extracted from the {\it Suzaku} public archive were used in the T13 analysis. These selection criteria yielded a sample of 76 observations of 43 objects: 24 objects were classified as type 1-1.2, 16 objects were classified as type 1.5, and 3 objects were classified as type 1.8-1.9. Sources with multiple observations were reduced and analyzed individually. 

T13 extracted the 15-50 keV and 2-10 keV energy flux ratio and Fe K$\alpha$ equivalent width (EW, calculated against the total continuum) for each observation, to characterize the gross spectral properties of the sample, and concluded that the joint consideration of the spectral hardness and EW of Fe K$\alpha$ emission for the local type 1 AGN population in their sample, and the deep, sharp Fe K absorption edges present in the hardest sample objects 
were consistent with reprocessing  by a quasi-spherical distribution of Compton-thick (N$_H$ $>$ 10$^{24}$ cm$^{-2}$) clouds. Simple considerations suggested that the absorbers are likely located within the broad-line region (BLR).  

In this paper, we investigate the role of the Compton-thick absorber across the entire local radio-quiet AGN population.  Specifically, we test our  Monte Carlo Radiative Transfer (MCRT) absorption model against the sample, to determine whether 
such a model is able to account for the properties of the local AGN population. 

\section{The Sample Selection}

We expand the sample set of T13 by adding the type 2 AGN in the 58-month BAT catalog that have available observations in the public domain 
of {\it Suzaku}. Following  the selection criteria of T13, we applied a cut-off for the archived data  of 2011 December 20 and a redshift limit of z $\lesssim$ 0.1,  we restricted our study to radio quiet objects and excluded LINERS.  This sample expansion yielded an additional 43 observations of 28 objects compared to the T13 sample. Details concerning the observations and data are presented in Table~\ref{tab:table}. We note that T13 omitted NGC 4138 from their analysis. The omission of this source does not change the conclusions of T13 as the source is within the average of the spectral properties in the T13 sample. We have added NGC 4138 in our study for completeness.

 \section{The Observational Data}
  \label{sec:sources}
{\em Suzaku} has four X-ray telescopes, each containing a silicon CCD within its focal plane forming the X-ray Imaging Spectrometers (XIS) suite. XIS0, XIS2 and XIS3 are front-illuminated (FI), providing data over a usable range of 0.6-10.0 keV with an energy resolution   FWHM  $\sim$ 130 eV at 6.0 keV. In November 2006, a charge leak was discovered in XIS2, making XIS0 and XIS3 the only operational FI chips. XIS1 is back-illuminated. The back-illuminated configuration extends the soft band to $\sim$0.2 keV; however, it also results in a lower effective area and higher background rate in the Fe K regime, compared to the FI chips. Consequently, XIS1 is excluded from our spectral analysis. {\it Suzaku} also carries the Hard X-ray Detector (HXD) that contains a silicon PIN diode detector covering a range of $10 - 100$ keV with a usable energy range of 15--50 keV. 

The data were reduced using HEAsoft v.6.10. The XIS cleaned event files were screened in XSELECT to exclude data  during passage through the South Atlantic Anomaly and also  excluding data starting 500 s before entry  and up to 500 s after exit. In addition, we excluded data having  an Earth elevation angle $<$ 10$^{\circ}$ and a cut-off rigidity $>$ 6 GeV.  CCDs were in 3 x 3 and 5 x 5 edit modes, with normal clocking mode. Good events were selected, having  grades 0, 2, 3, 4, and 6, while  hot and flickering pixels were removed using the SISCLEAN script. The spaced-row charge injection was utilized. XIS spectra were extracted from circular regions of 3.0$\arcmin$ radius centered on the source, while the background was extracted from a region of the same size offset from the source and from the corners of the chip that register calibration data. 

The cleaned PIN data were reduced utilizing the ftool {\sc hxdpinxbpi}. This tool calculates good time intervals (GTIs) of the non X-ray instrumental background (NXB, using model 'D' released 2008 June 17\footnote{http://www.astro.isas.jaxa.jp/suzaku/doc/suzakumemo/suzakumemo-2007-01.pdf} ) data that overlap with the source data and extracts both the source spectrum and NXB background spectrum through that common GTI. A simulated cosmic X-ray background (CXB) spectrum is calculated, based on the CXB spectrum reported in \citet{Boldt:1987kx} and renormalized to the 35$^\prime$ $\times$ 35$^\prime$ field of view of the PIN instrument. This is then combined with the NXB spectrum to produce a total PIN background spectrum for the observation. A dead-time correction typically 4-5\% is  applied to the source spectrum using the ftool {\sc hxddtcor}. Finally, the PIN data have a known 1$\sigma$ systematic uncertainty of 1.3\%\footnote{http://heasarc.nasa.gov/docs/suzaku/analysis/watchout.html}, which was applied to the PIN data as a quadrature sum of the systematic error and the statistical error using {\sc grppha}.  

During spectral fitting, we scaled the PIN part of the model by an energy-independent constant, as appropriate for the XIS or HXD nominal aim point used, 1.16 or 1.18, respectively, with the exception of Fairall 9 OSBID 705063010. Due to calibration issues during that observation of Fairall 9, we used an energy-independent constant of 1.30 ({c.f. \citealt{Lohfink:2012fk}). The scaling factors accounts for the cross-calibration constant required to correctly reconcile  the XIS and PIN data based on the calibration of the Crab Nebula \citep{Maeda:2008ab}.  

\section{Spectral Analysis and Sample Results}
 \label{sec:fitting}
 
\subsection{Estimation of Fe\,K$\alpha$ equivalent width and spectral hardness ratio}

To estimate the line equivalent widths and spectral hardness ratios, 
we fitted the 2-50 keV bandpass of each observation with simple phenomenological models, using {\sc XSPEC} v 12.5. The model used was a powerlaw partially-covered by a neutral Compton-thick absorber, parameterized by using the {\sc tbabs} model \citep{Wilms:2000qy}, fixing the column density to N$_H$ = 2 $\times$ 10$^{24}$ cm$^{-2}$.  We allowed a Gaussian emission component at $\sim$ 6.4 keV fit with a freely varying width ($\sigma$), normalization and energy. 
We included a full-covering column screen of gas representing the Galactic line-of-sight absorption, parameterized using   {\sc tbabs}  \citep{Wilms:2000qy} and fixed to the weighted average N$_H$ in the Dickey and Lockman survey \citep{Dickey:1990uq}.   This model was not intended to provide a physically meaningful description of the data but rather a simple parameterization from which fluxes can be measured for the different wavebands of interest. 

Following T13, we then extracted the total observed energy fluxes (ergs cm$^{-2}$ s$^{-1}$) for the 2-10 keV and 15-50 keV bandpasses for the type 2 AGN in the sample to determine the hardness ratio, Flux$_{15-50\, \textnormal{keV}}$/Flux$_{2-10\, \textnormal{keV}}$, for each observation. Hereafter, all fluxes and luminosities are the observed values, unless otherwise noted. These bandpasses were chosen as the most meaningful and practical for determination of the hardness of the X-ray spectrum associated with the active nucleus. In this field, the 2-10 keV band has become a standard bandpass over which to quote a flux, this is partly because many previous X-ray detectors covered this bandpass (e.g. the {\it EXOSAT} ME, {\it ASCA} SIS and GIS, {\it XMM} EPIC CCDs and the 
{\it Suzaku} XISs). Prior to {\it NuSTAR} there have been no reliable data available between $10 -15$ keV. On the high end, the upper limit was chosen based on the limited sensitivity of the PIN above 50 keV.   In determining the energy-band fluxes, the Galactic absorption had a negligible effect on the $2 - 10$ keV and 
$15 - 50$ keV flux measurements for the entire sample. Extending on the work of T13, we fit the $0.5 - 2$ keV bandpass of each observation with a powerlaw plus multiple Gaussian emission components, where required. We extracted the $0.5 - 2$ keV energy flux for the sample and corrected for Galactic absorption. The Galactic absorption had a non-negligible effect on the $0.5 - 2$ keV flux.  

Following T13, for each observation in our sample, we plotted the Flux$_{15 - 50\, \textnormal{keV}}$/Flux$_{2-10\, \textnormal{keV}}$ against the BAT flux (Figure~\ref{fig:Hardness}). The lower and upper solid black lines are the weighted mean Flux$_{15-50\, \textnormal{keV}}$/Flux$_{2-10\, \textnormal{keV}}$ for the type 1 ($1.73$) with variance  0.42 and type 2 AGNs   ($ 10.02$) with variance 22.75, respectively. We found the weighted mean hardness ratios by fitting the type 1 and type 2 AGN datasets with a constant model line.
The hardness ratio distribution of the type 2 AGN overlaps with the type 1 AGN distribution, suggesting a continuous distribution of spectral hardness across the AGN population (Figure~\ref{fig:Hardness}).

T13 also plotted the Flux$_{15-50\, \textnormal{keV}}$/Flux$_{2-10\, \textnormal{keV}}$ against the EW of the total Fe K$\alpha$ emission line  (measured against the total observed continuum). The addition of the type 2 AGN is shown in Figure~\ref{fig:EW}, it reveals a continuous distribution of spectral hardness and line EW across the AGN population. 

\subsection{Comparison to {\it BeppoSAX}}

It is instructive to compare the above results with those obtained
previously from  {\it BeppoSAX} observations.
{\it BeppoSAX} \citep{Boella:1997uq} was the first X-ray mission to cover three orders of magnitude in energy (0.1--300 keV) and comprised the Medium Energy Concentrator Spectrometer and the Photoswich Detection System, covering the 1.3--10 keV and 15--300 keV bandpasses, respectively. The broad bandpass of {\it BeppoSAX} allowed for simultaneous soft, medium, and hard X-ray data. 

Utilizing this broadband capability, \citet[hereafter D07]{Dadina:2007fk} and \citet[hereafter D08]{Dadina:2008lr} conducted an archival study of the Seyfert galaxies (type 1-1.9 and type 2) at z $\leq$ 0.1 observed with {\it BeppoSAX}. The sample overlaps  our own,  although we note that the selection criteria were different: D07 did not select the sources on hard X-ray flux and did allow inclusion of some radio-loud AGN.   D07 fitted these sources with a template model consisting of a powerlaw, a gaussian component  and a reflection component, with more complex components added when statistically required. During spectral fitting, D07 allowed the photon index to float, finding some relatively flat indices in the sample. D07 also allowed the Fe K$\alpha$ emission line and the reflection continuum to be modeled independently.  D08 statistically analyzed the D07 results to understand the characteristics of the nearby Seyfert population and found R = 1.23 $\pm$ 0.11, suggesting that the hard excess phenomenon is not ubiquitous in AGN, in contrast to the results of \citet{Tatum:2013lr}.  

To understand this discrepancy, we cross-correlated the T13 sample with the D07 sample and calculated the ratio Flux$_{15-50\,\textnormal{keV}}$/Flux$_{2-10 \,\textnormal{keV}}$ for the overlapping sources and overlaid them on our {\it Suzaku} sample. The {\it BeppoSAX} data show 
very hard X-ray spectral forms for the D07 sample (Figure~\ref{fig:Hardness}) with a mean hardness ratio 1.87.  While our sample and that of D07 span the local AGN population, it is not surprising that our sample shows a  greater number of hard sources, as our sample was hard X-ray selected.

As a more stringent  test of our hardness result we compared the IC4329A spectra 
from {\it BeppoSAX} (July 21, 1998), {\it Suzaku} (August 1, 2007) and {\it NuSTAR} (August 12, 2012) observations. {\it NuSTAR} is the first hard X-ray imaging telescope (FWHM angular resolution $\sim$ 17$\arcsec$), covering the 5-80 keV bandpass. The focusing optics and CZT detectors allow for low background rates and sensitivity two orders of magnitude better than previous hard X-ray missions \citep{Harrison:2010kx}. As shown in Figure~\ref{fig:Beppo_Suzaku}, the spectral shape is consistent among all the observations, showing the same result is obtained from all broad-band detectors.

If we apply the spectral models of D07 to our {\it Suzaku} spectra, we get broad agreement with D07. However, we believe our modeling approach is well justified. Large sample studies such as that of \citet{scott11a} have found $\Gamma=1.99\pm 0.01 $,  in agreement with detailed analysis of high quality data for some well-studied objects  \citep[e.g. $\Gamma \sim$ 2.1 for MCG-06-30-15][]{Miller:2008kx}. Such an analysis involved inclusion of a complex, multi-zoned absorber into the spectral model.   Continuum slopes around $\Gamma \sim 2$ are steeper than those fitted by D07, and if such an index was fixed in the D07 fits then 
the reflection contributions returned by the {\it BeppoSAX} data would necessarily be higher than those tabulated by D07. The D07 modeling approach returned  not only flat photon indices, but also required the photon index to vary over a large range of values ($\Delta \Gamma \sim 0.3$) in some individual sources, on timescales of weeks to months.  Such drastic changes in spectral shape within observations of an individual object may be explained better by variations in the absorber along the line of sight  rather  than large changes in the continuum form \citep[e.g.][]{Miller:2008kx}. 
We conclude that the  lack of any strong hard components in the fit results from the D07 analysis is attributable to the different approach used for spectral fitting.  We move forward by analyzing the data in the context of our preferred model.

\section{Monte-Carlo Radiative Transfer Calculation}

Our aim in this paper is to test whether we may explain the gross X-ray 
spectral properties of the local AGN population with a simple, but physically self-consistent, model of
reprocessing by circumnuclear material. To this end we use a simple 
Monte-Carlo Radiative Transfer (MCRT) code, which was originally presented in T13. 
We present here a general overview and refer the reader to T13 for specific details. 

The MCRT calculation presented in T13 uses a simple model of a partial covering atmosphere. At every point this atmosphere is assumed to be a strict two-phase medium which is either vacuum or cold, neutral gas at constant density. In reality the atmosphere is likely to be hot and ionized, however to explore a wide range of parameter space, the model presented here makes the simplifying assumption of cold gas clouds, such that all ions heavier than H are neutral, but with H ionized, so that the number of free scattering electrons equals the number of H atoms. An inhomogeneous model atmosphere is created by randomly placing a set of spherical gas clouds within a spherical annulus centered on the primary source, with  the annulus having a defined inner and outer radius. The results are independent of any scale size for the cloud distribution, as all length scales are relative to an individual sphere's radius. The probability density distribution from which these points are drawn is symmetric azimuthally, but may have some dependence on polar angle. If two (or more) spheres overlap, the density is not doubled, but remains at the same value as elsewhere in the gas phase. For distributions with a high density of points, the spheres heavily overlap, and the resulting gas distribution does not then resemble a set of individual blobs, but rather a sponge-like topology. It is assumed that the intrinsic X-ray continuum source is relatively compact and likely associated with an accretion disk corona. For the radiative transfer calculation, the spherical distribution is assumed to be bisected by an infinite, thin accretion disk, that absorbs photons but does not itself radiate at X-ray energies, so that the scattered light received by the observer is reduced by a factor two compared with the optically-thin case without the disk.  The reason for setting the accretion disc as being a black absorber is to show that any results we obtain are due entirely to the cloud distribution.  We might also expect some disc reflection, and more sophisticated modeling could include that. However, the main objective in using MCRT is not to carry out a detailed and realistic modeling of X-ray spectra, but rather to determine if a simple model that nonetheless has some self-consistent radiative transfer physics can explain the broad spectral properties of the entire AGN population. 
 
As photons propagate through the distribution they may be either absorbed by the gas, or be Compton scattered.  After absorption, there may be resulting Fe K fluorescent line emission.  This is implemented by calculating the mean free path for the photon in the gas due to three processes: 
\begin{enumerate}
\item Compton scattering, using the full energy-dependent Klein-Nishina cross-section, integrated over all scattering angles.
\item Absorption by elements other than K-shell transitions of Fe, in which case the photon disappears from the calculation (i.e. line emission from elements other than Fe K-shell transitions, primarily in the soft X-ray band, is neglected).
\item Absorption by an Fe K-shell transition, which may be followed by Fe\,K$\alpha$ or Fe\,K$\beta$ fluorescent line emission.
\end{enumerate}

\subsection{Modifications to MCRT}
The MCRT code used in T13 generated spectra by summing the observed photons over a large number of sightlines, effectively producing a spectrum  averaged over many sightlines for a given input atmosphere. However, one expects considerable variations between spectra generated for different sightlines within the same atmosphere, considering the highly inhomogeneous nature of this partial covering model, and in this paper we investigate whether the dispersion in the observed properties of AGN may be attributed to natural variation between differing sightlines. In order to sample the range of sightlines produced by a single atmosphere, this code was modified to generate spectra binned over a full range of both azimuthal and polar angles. Taken together, they covered the full 2$\pi$ sr. The hemisphere was divided into regions of equal solid angle, each covering a small range in both the polar angle ($\theta$) and azimuthal angle ($\phi$), and each corresponding to a potential observed sightline. 

In each spectrum, an effective observed solid angle was determined by requiring it to be sufficiently large such that the majority of sightlines are partially covered and sufficiently small to produce variation between sightlines. In choosing the solid angle of a sightline, we effectively estimate the ratio between the characteristic size of the atmosphere and the size of the X-ray emitting region. We expect the scale of the source and the cloud structure might be similar \citep{Turner:2009lr}; therefore, we divided the polar angle into bin sizes equaling the angle subtended by one radius at the inner edge of the spherical distribution, approximately 10$^o$. Each bin was then subdivided into azimuthal bins with the same size criterion. 

In order to investigate the effects of anisotropy in the AGN atmosphere, we further modified the code to allow the clouds to be placed with a spatial density that is either constant, proportional to sin($\theta$), or proportional to sin$^2$($\theta$), allowing the potential for the cloud density to be preferentially aligned along the plane of the accretion disk. Figure~\ref{fig:sinpower} displays the results of three such simulations. Note that only the number of clouds per unit volume changed, while the density of the gas remained constant. The hardness ratio and equivalent width distribution has a dependence on geometry. The HR-EW relationship for constant cloud density is consistent with any polar viewing angle. However, as the cloud distribution becomes more anisotropic, the higher equivalent widths and hardness ratios correspond to larger polar viewing angles, i.e., more edge-on sightlines.  

Note that the MCRT model results are invariant with respect to any choice of electron density or cloud size, these parameters may be subsumed into a single model parameter, which is the column density through the diameter of one cloud.  The results are also invariant with respect to the size of overall the cloud distribution, for a given column density and cloud geometry, the whole distribution may be scaled up or down in size without affecting the resulting spectra.

\section{Interpretation of the distribution of results in the hardness ratio - Fe K$\alpha$ plane  \label{mcrtanalysis}}

We compared our observational result with calculations from the modified MCRT code. The calculations in Figure~\ref{fig:EW} were performed assuming 1000 clouds in the spherical distribution, an inner radius of 10 units, an outer radius 20 units, $\Gamma$=2.0, a sin($\theta$) cloud distribution and N$_H$ = 9 x 10$^{23}$ and 2 x 10$^{24}$ cm$^{-2}$. These simulations are consistent with the type 1 and type 2 spectral properties of our sample. In the N$_H$=2 x 10$^{24}$ cm$^{-2}$ simulation, 48\% and 52\% of sightlines were obscured by Compton-thick (N$_H$ $>$ 10$^{24}$ cm$^{-2}$) and Compton-thin (N$_H$ $<$ 10$^{24}$ cm$^{-2}$) columns of gas, respectively.  

As shown in Figure~\ref{fig:EW}, the modified MCRT model is capable of creating a distribution of spectra that follows the general trend visible in the observational data. However, in order to make a meaningful comparison between the two, we require a statistical procedure than can quantify the similarity between the two distributions. In this case, we used the 1D Kolmogorov-Smirnoff (KS) test and treated the MCRT model as the probability distribution and the observational data as the sample. 

A useful feature of the KS test is that the distribution of the test statistic is independent of the probability distribution function being tested. This allows us to perform it relatively easily, without needing to determine what the distribution of the test statistic should be for each set of model results individually. 
A disadvantage of the test, however, is that the statistical significance of the test statistic is unknown if model
parameters are varied.  In this paper, our aim is not to carry out a full statistical test of goodness-of-fit, as the
model is certainly too simplistic, in that it assumes that every AGN in the sample has an identical distribution of
circumnuclear gas, with only the line of sight to the nucleus differing between AGN.  Hence we use the KS test 
only as indicative of the relative success of our choice of basic models.
We performed two 1D KS tests, treating the EW and HR independently of each other. Although this approach loses information about the correlation in the data, it remains helpful as an indicator of goodness of fit. Before the model could be tested against the data, it was important to consider the experimental errors present in the observations. These errors are vastly greater than the Monte Carlo noise in the MCRT model spectra. If the model were a good description of the X-ray spectra of AGN, we would expect to observe that distribution of spectra blurred by considerable experimental errors. To meaningfully compare model and observation, we must generate the distribution of HRs and EWs that we would expect to recover with our instruments if we were to observe the spectra generated by the MCRT model. Therefore, before any KS tests were performed it was necessary to add simulated experimental errors to the model's predictions.

One way to generate simulated experimental errors is to make use of the errors present in the observational data. Because we are treating the EW and HR as independent for the purposes of our statistical tests, a reasonable way to generate an error for the EW of a model point is to select the {\it n} closest observational data points along the EW axis and assign the model point a standard error equal to the average of the standard errors on those {\it n} data points, likewise for the HR. A blurred model distribution can then be produced by randomly generating a number of new points for each model point, with the new points drawn from a normal distribution centered on the original model point and with a standard deviation equal to the error which we have assigned to that point.

In order to carry out the blurring with simulated experimental errors and the pair of KS tests, we generated 1000 scattered points for each original model point, with errors derived by averaging those of the nearest 10 observational data points in logarithmic space. This resulted in error-scattered distributions with low Monte Carlo noise, which reduced to approximately continuous cumulative density functions along both EW and HR axes (a requirement of the KS testing process). We then performed a pair of 1D KS tests treating the two variables as independent. An example of a model distribution which has been scattered in this way can be seen in Figure~\ref{fig:errors}.

The results of performing the KS tests were dependent on the assumed input parameters, with particularly high dependence on the column density per cloud (N$_H$). We obtained the highest p-values, p$_{EW}$ = 0.75 and p$_{HR}$ = 0.046 for the EW and HR 1D KS tests, assuming N$_H$ = 10$^{24}$ cm$^{-2}$, $\Gamma$ = 2.1, cloud density distribution proportional to sin$\theta$, and outer radius= 20 cloud radii. 

With values of up to 0.75 for p$_{EW}$, the MCRT model appears to generate a very similar distribution of EWs to that observed in both type 1 and type 2 Seyfert galaxies. Examination of 
Figures~\ref{fig:EW} and ~\ref{fig:errors}  shows the model predictions include a  tail of very low hardness ratios  that represent 
the least obscured sight-lines, where a direct view to the source results in no absorption of soft X-rays and hence no excess hardness in the spectrum. 
Indeed, several so-called 'bare' Seyfert galaxies have been reported in the literature (e.g. Ark 120 and Fairall-9, \citealt{patrick11a}). Overall,  the distribution of simulated spectra and those in our observed population are consistent, suggests that partial covering atmospheres of Compton thick gas may play an important part in the unified model of AGN X-ray spectra.

\section{X-ray Luminosity Correlation}

Next we  investigated the relationship between the luminosities in different X-ray bands, with the goal of improving our understanding of the reprocessor. 
Figure~\ref{fig:lum_lum} shows the 15--50 keV / 2-10 keV luminosity ratio and the 15-50/0.5-2 keV luminosity ratio plotted against both the 15-50 keV luminosity and 0.5-2 keV luminosity. 
It is evident (top panels) that the type 2 AGN have the same hard-band (15-50 keV) luminosity distribution as the rest of the AGN population. This supports the 
expectation, from Unified Models, that the 
different optical classes should harbor nuclei of the same power.  The uniformity of luminosity distributions for type 1 and type 2 AGN is indicative that the 15-50 keV luminosity is a fair representation of the true intrinsic continuum luminosity. 

It is also clear that the type 2 AGN (aqua stars) have systematically lower soft-band luminosities than the rest of the population (lower panels): this is already well known and thought to be due to the relatively high obscuration of the primary continuum  for type 2 AGN versus type 1 AGN in this soft energy band \citep{halderson01}.  A few intermediate type AGN appear at low soft-band luminosities, as expected from our cloud model, where the obscuration is a probability function across the AGN optical types and also varies with time for an individual object. 

The type 2 AGN show a broader distribution of  hardness ratios relative to  type 1 sources, again, consistent with our view that the absorber is clumpy and has a polar angle dependence for the cloud distribution.  Type 2 AGN are viewed through the largest number of clouds (`edge-on to the distribution') and can therefore produce the hardest spectra.  However, the Type 2's also have the greatest potential to exhibit  a large range of hardness,  as these clouds move in and out of the line-of-sight. 

\section{Discussion} 

We have extended the work of T13 to consider the broad properties of a hard X-ray selected set of local AGN, thought to provide an unbiased representation of the local population. The consistency 
of the luminosity distributions of type 1 and type 2 AGN in the 15-50 keV band indicates that this band does indeed provide a relatively unbiased selection criterion. 
The spectral hardness and Fe K$\alpha$ have once again been used as a diagnostic of the reprocessing gas. Extending the results of T13, we have found that the local radio-quiet AGN population yields a continuous distribution of values for spectral hardness and line equivalent width. A detailed comparison shows that our {\it Suzaku} results are consistent with measurements across a similar bandpass from previous observations using {\it BeppoSAX}.  

The range of X-ray spectral properties exhibited across the population can be parameterized within the context of a simple circumnuclear cloud model, calculated 
here using Monte Carlo radiative transfer techniques. 
As proof of the concept, we have fitted the sample with a model atmosphere that has a column density $N_H = 10^{24} {\rm atoms\, cm^{-2}}$ with 
inner and outer radii for the shell set at 10 and 20 cloud radii, and using 1000 clouds filling our shell with a sin$\theta$ dependence of the cloud density on polar angle. The range of observed AGN types are then explained primarily by different sight-lines through the cloud ensemble to the nucleus. 
This shows that the obscuring material required by unified theories may be related to the X-ray absorber seen in type 1 AGN, as previously suggested from X-ray studies \citep[e.g.][]{cappi00}.  

There is significant short-timescale variability in the X-ray absorption of both type 1 and type 2 AGN \citep[e.g.,][and references therein]{Turner:2000lr,  Miller:2008kx, risaliti2010a}, which, in the most extreme cases produces a so-called `changing look' AGN \citep[e.g.][]{matt2003a,braito2014a}. 
A clumpy X-ray absorber is required to explain the observed X-ray absorption variability of AGN, as also suggested by  \citet[e.g][]{markowitz2014a}. 
The characteristic variability timescales coupled with the result of this work supports previous suggestions that the X-ray absorber is part of a massive outflow, whose presence is observed across a broad bandpass in energy \cite[e.g.,][]{Elitzur:2006fk,elitzur08a}.    The clumpy nature of the torus is not only 
evidenced from X-ray observations, but  has also been suggested based on theoretical models  to explain the isotropic infrared emission coupled with the anisotropic obscuration of AGN \citep[e.g.][]{,nenkova08b}.  

 \cite{Ricci:2014lr} studied the relationship between the narrow Fe K$\alpha$ emission line and the 10-50 keV continuum in Seyfert 1 and 2s and found the L$_{Fe K\alpha}$/L$_{10-50\,\textnormal{keV}}$ is lower on average in Seyfert 2s than Seyfert 1s,  consistent with the idea that  the Fe K line emission may be self-absorbed by the reprocessor in type 2 objects, which is evident in our MCRT model results.  
  
In the context of MCRT, T13 estimated the location of the Compton-thick, X-ray absorber to be within the optical broad line region. \citet[][hereafter S11]{Shu:2011lr} studied the Fe K alpha emission cores of 10 type 2 AGN (4 of which are included in our study) and compared them to the Fe K alpha emission core of 13 type 1 AGN. They found a marginal difference between the Fe K alpha luminosity in type 1 and type 2 AGN and concluded the Fe K alpha emitting region for the type 2 AGN lies between the outer part of the  optical broad line region and the putative torus.

\section{Conclusions}    

Using the 58-month BAT catalog, we have identified and analyzed an unbiased sample of {\it Suzaku} observations representative of the local population of radio-quiet AGN. We have demonstrated how a single, simple cloud reprocessing ensemble can explain the gross sample properties across all AGN types.

Examination of the spectral hardness and Fe K$\alpha$ equivalent width shows a continuous distribution of properties from type 1 to type 2 AGN. 
Comparison of the data with our Monte Carlo Radiative Transfer model for a reprocessing cloud ensemble shows the local population of radio-quiet AGN is  consistent with reprocessing of the X-ray continuum in an ensemble of  low-ionization, Compton-thick clouds.  This cloud distribution allows us to reproduce the entire range of observed properties simply by changing the observer's sightline. Of course, other cloud variations across the sample ($\xi$, N$_H$ etc) are expected, but the primary cause
of the correlated variation in hardness ratio and line equivalent width for the entire sample of type\,1 and type\,2 AGN
may be ascribed simply to differing lines of sight through a clumpy circumnuclear distribution of absorbing and Compton scattering gas.

Such an X-ray absorber would likely form part of a clumpy, massive outflow, that would shape the broad observed  spectral shape of the AGN, and would form an important component of material and energetic feedback between the galaxy nucleus and the host. 

\section*{Acknowledgements}

This research was supported by an appointment to the NASA Postdoctoral Program at the Goddard Space Flight Center, administered by Oak Ridge Associated Universities through a contract with NASA. TJT would like to acknowledge NASA grant  NNX11AJ57G. 

\bibliographystyle{apj}      
\bibliography{list} 

\begin{thebibliography}{44}
\expandafter\ifx\csname natexlab\endcsname\relax\def\natexlab#1{#1}\fi

\bibitem[{{Baumgartner} {et~al.}(2010){Baumgartner}, {Tueller}, {Markwardt}, \&
  {Skinner}}]{Baumgartner:2010fk}
{Baumgartner}, W.~H., {Tueller}, J., {Markwardt}, C., \& {Skinner}, G. 2010, in
  Bulletin of the American Astronomical Society, Vol.~42, AAS/High Energy
  Astrophysics Division \#11, 675

\bibitem[{{Boella} {et~al.}(1997){Boella}, {Butler}, {Perola}, {Piro},
  {Scarsi}, \& {Bleeker}}]{Boella:1997uq}
{Boella}, G., {Butler}, R.~C., {Perola}, G.~C., {Piro}, L., {Scarsi}, L., \&
  {Bleeker}, J.~A.~M. 1997, \aaps, 122, 299

\bibitem[{{Boldt}(1987)}]{Boldt:1987kx}
{Boldt}, E. 1987, in IAU Symposium, Vol. 124, Observational Cosmology, ed.
  A.~{Hewitt}, G.~{Burbidge}, \& L.~Z. {Fang}, 611--615

\bibitem[{{Braito} {et~al.}(2014){Braito}, {Reeves}, {Gofford}, {Nardini},
  {Porquet}, \& {Risaliti}}]{braito2014a}
{Braito}, V., {Reeves}, J.~N., {Gofford}, J., {Nardini}, E., {Porquet}, D., \&
  {Risaliti}, G. 2014, \apj, 795, 87

\bibitem[{{Cappi} {et~al.}(2000){Cappi}, {Bassani}, {Malaguti}, {Palumbo},
  {Dadina}, {Comastri}, {Di Cocco}, {Blanco}, {dal Fiume}, {Fabian},
  {Frontera}, {Guainazzi}, {Maccacaro}, {Maiolino}, {Matt}, {Piro},
  {Trifoglio}, \& {Zhang}}]{cappi00}
{Cappi}, M., {Bassani}, L., {Malaguti}, G., {Palumbo}, G.~G.~C., {Dadina}, M.,
  {Comastri}, A., {Di Cocco}, G., {Blanco}, P., {dal Fiume}, D., {Fabian}, A.,
  {Frontera}, F., {Guainazzi}, M., {Maccacaro}, T., {Maiolino}, R., {Matt}, G.,
  {Piro}, L., {Trifoglio}, M., \& {Zhang}, N. 2000, Advances in Space Research,
  25, 815

\bibitem[{{Crenshaw} {et~al.}(2003){Crenshaw}, {Kraemer}, \&
  {George}}]{Crenshaw:2003vn}
{Crenshaw}, D.~M., {Kraemer}, S.~B., \& {George}, I.~M. 2003, \araa, 41, 117

\bibitem[{{Dadina}(2007)}]{Dadina:2007fk}
{Dadina}, M. 2007, \aap, 461, 1209

\bibitem[{{Dadina}(2008)}]{Dadina:2008lr}
---. 2008, \aap, 485, 417

\bibitem[{{Dickey} \& {Lockman}(1990)}]{Dickey:1990uq}
{Dickey}, J.~M. \& {Lockman}, F.~J. 1990, \araa, 28, 215

\bibitem[{{Elitzur}(2008)}]{elitzur08a}
{Elitzur}, M. 2008, \nat, 52, 274

\bibitem[{{Elitzur} \& {Shlosman}(2006)}]{Elitzur:2006fk}
{Elitzur}, M. \& {Shlosman}, I. 2006, \apjl, 648, L101

\bibitem[{{Fabian}(1994)}]{Fabian:1994gf}
{Fabian}, A.~C. 1994, \apjs, 92, 555

\bibitem[{{Gofford} {et~al.}(2013){Gofford}, {Reeves}, {Tombesi}, {Braito},
  {Turner}, {Miller}, \& {Cappi}}]{Gofford:2013lr}
{Gofford}, J., {Reeves}, J.~N., {Tombesi}, F., {Braito}, V., {Turner}, T.~J.,
  {Miller}, L., \& {Cappi}, M. 2013, \mnras, 430, 60

\bibitem[{{Halderson} {et~al.}(2001){Halderson}, {Moran}, {Filippenko}, \&
  {Ho}}]{halderson01}
{Halderson}, E.~L., {Moran}, E.~C., {Filippenko}, A.~V., \& {Ho}, L.~C. 2001,
  \aj, 122, 637

\bibitem[{{Harrison} {et~al.}(2010){Harrison}, {Boggs}, {Christensen}, {Craig},
  {Hailey}, {Stern}, {Zhang}, {Angelini}, {An}, {Bhalerao}, {Brejnholt},
  {Cominsky}, {Cook}, {Doll}, {Giommi}, {Grefenstette}, {Hornstrup}, {Kaspi},
  {Kim}, {Kitaguchi}, {Koglin}, {Liebe}, {Madejski}, {Kruse Madsen}, {Mao},
  {Meier}, {Miyasaka}, {Mori}, {Perri}, {Pivovaroff}, {Puccetti}, {Rana}, \&
  {Zoglauer}}]{Harrison:2010kx}
{Harrison}, F.~A., {Boggs}, S., {Christensen}, F., {Craig}, W., {Hailey}, C.,
  {Stern}, D., {Zhang}, W., {Angelini}, L., {An}, H., {Bhalerao}, V.,
  {Brejnholt}, N., {Cominsky}, L., {Cook}, W.~R., {Doll}, M., {Giommi}, P.,
  {Grefenstette}, B., {Hornstrup}, A., {Kaspi}, V., {Kim}, Y., {Kitaguchi}, T.,
  {Koglin}, J., {Liebe}, C.~C., {Madejski}, G., {Kruse Madsen}, K., {Mao}, P.,
  {Meier}, D., {Miyasaka}, H., {Mori}, K., {Perri}, M., {Pivovaroff}, M.,
  {Puccetti}, S., {Rana}, V., \& {Zoglauer}, A. 2010, in Society of
  Photo-Optical Instrumentation Engineers (SPIE) Conference Series, Vol. 7732,
  Society of Photo-Optical Instrumentation Engineers (SPIE) Conference Series

\bibitem[{{Kaspi} {et~al.}(2002){Kaspi}, {Brandt}, {George}, {Netzer},
  {Crenshaw}, {Gabel}, {Hamann}, {Kaiser}, {Koratkar}, {Kraemer}, {Kriss},
  {Mathur}, {Mushotzky}, {Nandra}, {Peterson}, {Shields}, {Turner}, \&
  {Zheng}}]{Kaspi:2002lr}
{Kaspi}, S., {Brandt}, W.~N., {George}, I.~M., {Netzer}, H., {Crenshaw}, D.~M.,
  {Gabel}, J.~R., {Hamann}, F.~W., {Kaiser}, M.~E., {Koratkar}, A., {Kraemer},
  S.~B., {Kriss}, G.~A., {Mathur}, S., {Mushotzky}, R.~F., {Nandra}, K.,
  {Peterson}, B.~M., {Shields}, J.~C., {Turner}, T.~J., \& {Zheng}, W. 2002,
  \apj, 574, 643

\bibitem[{{Lohfink} {et~al.}(2012){Lohfink}, {Reynolds}, {Miller}, {Brenneman},
  {Mushotzky}, {Nowak}, \& {Fabian}}]{Lohfink:2012fk}
{Lohfink}, A.~M., {Reynolds}, C.~S., {Miller}, J.~M., {Brenneman}, L.~W.,
  {Mushotzky}, R.~F., {Nowak}, M.~A., \& {Fabian}, A.~C. 2012, \apj, 758, 67

\bibitem[{{Maeda} {et~al.}(2008){Maeda}, {Someya}, {Ishida}, {the XRT team},
  {Hayashida}, {Mori}, \& the XIS~team}]{Maeda:2008ab}
{Maeda}, Y., {Someya}, K., {Ishida}, M., {the XRT team}, {Hayashida}, K.,
  {Mori}, H., \& the XIS~team. 2008, JX-ISAS-SUZAKU-MEMO-2008-06

\bibitem[{{Markowitz} {et~al.}(2014){Markowitz}, {Krumpe}, \&
  {Nikutta}}]{markowitz2014a}
{Markowitz}, A.~G., {Krumpe}, M., \& {Nikutta}, R. 2014, \mnras, 439, 1403

\bibitem[{{Matt} {et~al.}(2003){Matt}, {Guainazzi}, \& {Maiolino}}]{matt2003a}
{Matt}, G., {Guainazzi}, M., \& {Maiolino}, R. 2003, \mnras, 342, 422

\bibitem[{{Miller} \& {Turner}(2013)}]{Miller:2013fk}
{Miller}, L. \& {Turner}, T.~J. 2013, \apj, 773, L5

\bibitem[{{Miller} {et~al.}(2008){Miller}, {Turner}, \&
  {Reeves}}]{Miller:2008kx}
{Miller}, L., {Turner}, T.~J., \& {Reeves}, J.~N. 2008, \aap, 483, 437

\bibitem[{{Miller} {et~al.}(2007){Miller}, {Turner}, {Reeves}, {George},
  {Kraemer}, \& {Wingert}}]{Miller:2007fe}
{Miller}, L., {Turner}, T.~J., {Reeves}, J.~N., {George}, I.~M., {Kraemer},
  S.~B., \& {Wingert}, B. 2007, \aap, 463, 131

\bibitem[{{Nenkova} {et~al.}(2008){Nenkova}, {Sirocky}, {Nikutta},
  {Ivezi{\'c}}, \& {Elitzur}}]{nenkova08b}
{Nenkova}, M., {Sirocky}, M.~M., {Nikutta}, R., {Ivezi{\'c}}, {\v Z}., \&
  {Elitzur}, M. 2008, \apj, 685, 160

\bibitem[{{Patrick} {et~al.}(2011){Patrick}, {Reeves}, {Porquet}, {Markowitz},
  {Lobban}, \& {Terashima}}]{patrick11a}
{Patrick}, A.~R., {Reeves}, J.~N., {Porquet}, D., {Markowitz}, A.~G., {Lobban},
  A.~P., \& {Terashima}, Y. 2011, \mnras, 411, 2353

\bibitem[{{Pounds} {et~al.}(2003){Pounds}, {Reeves}, {King}, {Page}, {O'Brien},
  \& {Turner}}]{Pounds:2003uq}
{Pounds}, K.~A., {Reeves}, J.~N., {King}, A.~R., {Page}, K.~L., {O'Brien},
  P.~T., \& {Turner}, M.~J.~L. 2003, \mnras, 345, 705

\bibitem[{{Rees}(1977)}]{rees77a}
{Rees}, M.~J. 1977, \qjras, 18, 429

\bibitem[{{Reeves} {et~al.}(2009){Reeves}, {O'Brien}, {Braito}, {Behar},
  {Miller}, {Turner}, {Fabian}, {Kaspi}, {Mushotzky}, \&
  {Ward}}]{Reeves:2009zt}
{Reeves}, J.~N., {O'Brien}, P.~T., {Braito}, V., {Behar}, E., {Miller}, L.,
  {Turner}, T.~J., {Fabian}, A.~C., {Kaspi}, S., {Mushotzky}, R., \& {Ward}, M.
  2009, \apj, 701, 493

\bibitem[{{Ricci} {et~al.}(2014){Ricci}, {Ueda}, {Paltani}, {Ichikawa},
  {Gandhi}, \& {Awaki}}]{Ricci:2014lr}
{Ricci}, C., {Ueda}, Y., {Paltani}, S., {Ichikawa}, K., {Gandhi}, P., \&
  {Awaki}, H. 2014, \mnras, 441, 3622

\bibitem[{{Risaliti} {et~al.}(2009){Risaliti}, {Braito}, {Laparola}, {Bianchi},
  {Elvis}, {Fabbiano}, {Maiolino}, {Matt}, {Reeves}, {Salvati}, \&
  {Wang}}]{Risaliti:2009lr}
{Risaliti}, G., {Braito}, V., {Laparola}, V., {Bianchi}, S., {Elvis}, M.,
  {Fabbiano}, G., {Maiolino}, R., {Matt}, G., {Reeves}, J., {Salvati}, M., \&
  {Wang}, J. 2009, \apjl, 705, L1

\bibitem[{{Risaliti} {et~al.}(2010){Risaliti}, {Elvis}, {Bianchi}, \&
  {Matt}}]{risaliti2010a}
{Risaliti}, G., {Elvis}, M., {Bianchi}, S., \& {Matt}, G. 2010, \mnras, 406,
  L20

\bibitem[{{Risaliti} {et~al.}(2013){Risaliti}, {Harrison}, {Madsen}, {Walton},
  {Boggs}, {Christensen}, {Craig}, {Grefenstette}, {Hailey}, {Nardini},
  {Stern}, \& {Zhang}}]{Risaliti:2013lr}
{Risaliti}, G., {Harrison}, F.~A., {Madsen}, K.~K., {Walton}, D.~J., {Boggs},
  S.~E., {Christensen}, F.~E., {Craig}, W.~W., {Grefenstette}, B.~W., {Hailey},
  C.~J., {Nardini}, E., {Stern}, D., \& {Zhang}, W.~W. 2013, \nat, 494, 449

\bibitem[{{Scott} {et~al.}(2011){Scott}, {Stewart}, {Mateos}, {Alexander},
  {Hutton}, \& {Ward}}]{scott11a}
{Scott}, A.~E., {Stewart}, G.~C., {Mateos}, S., {Alexander}, D.~M., {Hutton},
  S., \& {Ward}, M.~J. 2011, \mnras, 417, 992

\bibitem[{{Shu} {et~al.}(2011){Shu}, {Yaqoob}, \& {Wang}}]{Shu:2011lr}
{Shu}, X.~W., {Yaqoob}, T., \& {Wang}, J.~X. 2011, \apj, 738, 147

\bibitem[{{Tatum} {et~al.}(2013){Tatum}, {Turner}, {Miller}, \&
  {Reeves}}]{Tatum:2013lr}
{Tatum}, M.~M., {Turner}, T.~J., {Miller}, L., \& {Reeves}, J.~N. 2013, \apj,
  762, 80

\bibitem[{{Tatum} {et~al.}(2012){Tatum}, {Turner}, {Sim}, {Miller}, {Reeves},
  {Patrick}, \& {Long}}]{Tatum:2012fk}
{Tatum}, M.~M., {Turner}, T.~J., {Sim}, S.~A., {Miller}, L., {Reeves}, J.~N.,
  {Patrick}, A.~R., \& {Long}, K.~S. 2012, \apj, 752, 94

\bibitem[{{Terashima} {et~al.}(2009){Terashima}, {Gallo}, {Inoue}, {Markowitz},
  {Reeves}, {Anabuki}, {Fabian}, {Griffiths}, {Hayashida}, {Itoh}, {Kokubun},
  {Kubota}, {Miniutti}, {Takahashi}, {Yamauchi}, \& {Yonetoku}}]{terashima09a}
{Terashima}, Y., {Gallo}, L.~C., {Inoue}, H., {Markowitz}, A.~G., {Reeves},
  J.~N., {Anabuki}, N., {Fabian}, A.~C., {Griffiths}, R.~E., {Hayashida}, K.,
  {Itoh}, T., {Kokubun}, N., {Kubota}, A., {Miniutti}, G., {Takahashi}, T.,
  {Yamauchi}, M., \& {Yonetoku}, D. 2009, \pasj, 61, 299

\bibitem[{{Tombesi} {et~al.}(2010){Tombesi}, {Cappi}, {Reeves}, {Palumbo},
  {Yaqoob}, {Braito}, \& {Dadina}}]{Tombesi:2010yq}
{Tombesi}, F., {Cappi}, M., {Reeves}, J.~N., {Palumbo}, G.~G.~C., {Yaqoob}, T.,
  {Braito}, V., \& {Dadina}, M. 2010, \aap, 521, A57

\bibitem[{{Turner} \& {Miller}(2009)}]{Turner:2009lr}
{Turner}, T.~J. \& {Miller}, L. 2009, \aapr, 17, 47

\bibitem[{{Turner} {et~al.}(2009){Turner}, {Miller}, {Kraemer}, {Reeves}, \&
  {Pounds}}]{Turner:2009ys}
{Turner}, T.~J., {Miller}, L., {Kraemer}, S.~B., {Reeves}, J.~N., \& {Pounds},
  K.~A. 2009, \apj, 698, 99

\bibitem[{{Turner} {et~al.}(2000){Turner}, {Perola}, {Fiore}, {Matt}, {George},
  {Piro}, \& {Bassani}}]{Turner:2000lr}
{Turner}, T.~J., {Perola}, G.~C., {Fiore}, F., {Matt}, G., {George}, I.~M.,
  {Piro}, L., \& {Bassani}, L. 2000, \apj, 531, 245

\bibitem[{{Turner} {et~al.}(2008){Turner}, {Reeves}, {Kraemer}, \&
  {Miller}}]{Turner:2008qy}
{Turner}, T.~J., {Reeves}, J.~N., {Kraemer}, S.~B., \& {Miller}, L. 2008, \aap,
  483, 161

\bibitem[{{V{\'e}ron-Cetty} \& {V{\'e}ron}(2006)}]{veron06}
{V{\'e}ron-Cetty}, M.-P. \& {V{\'e}ron}, P. 2006, \aap, 455, 773

\bibitem[{{Wilms} {et~al.}(2000){Wilms}, {Allen}, \& {McCray}}]{Wilms:2000qy}
{Wilms}, J., {Allen}, A., \& {McCray}, R. 2000, \apj, 542, 914

\end{thebibliography}

\begin{figure}
\epsscale{.7}
\plotone{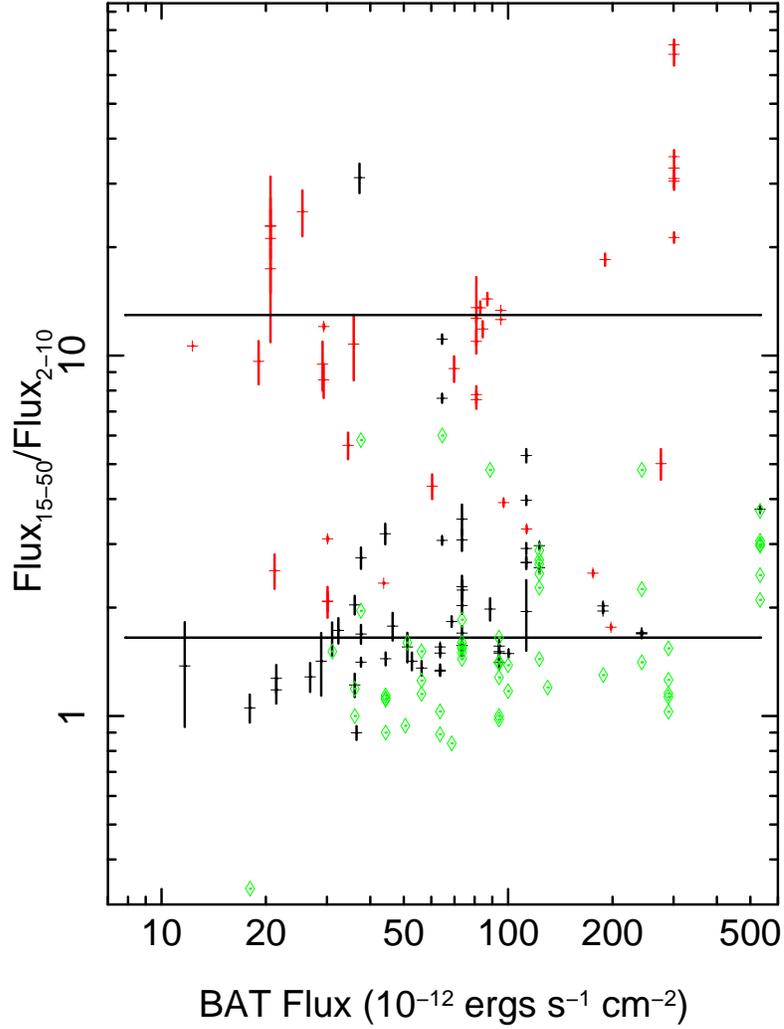}
\caption{The hardness ratio,  Flux$_{15-50\,\textnormal{keV}}$/Flux$_{2-10\,\textnormal{keV}}$, plotted against the {\it Swift} BAT flux, including both the type 1 (black markers) and type 2 (red markers) AGN observed using {\it Suzaku}. Overlaid are the hardness ratios we have extracted from the D07 sample points observed using {\it BeppoSax} (green). Solid lines represent the weighted hardness ratio mean of the type 1 (lower black line) and type 2 sources (upper black line) from the {\it Suzaku} sample analysis.}
\label{fig:Hardness}
\end{figure}

\begin{figure}
\centering
\scalebox{2}{\rotatebox{0}{\includegraphics[width=5cm]{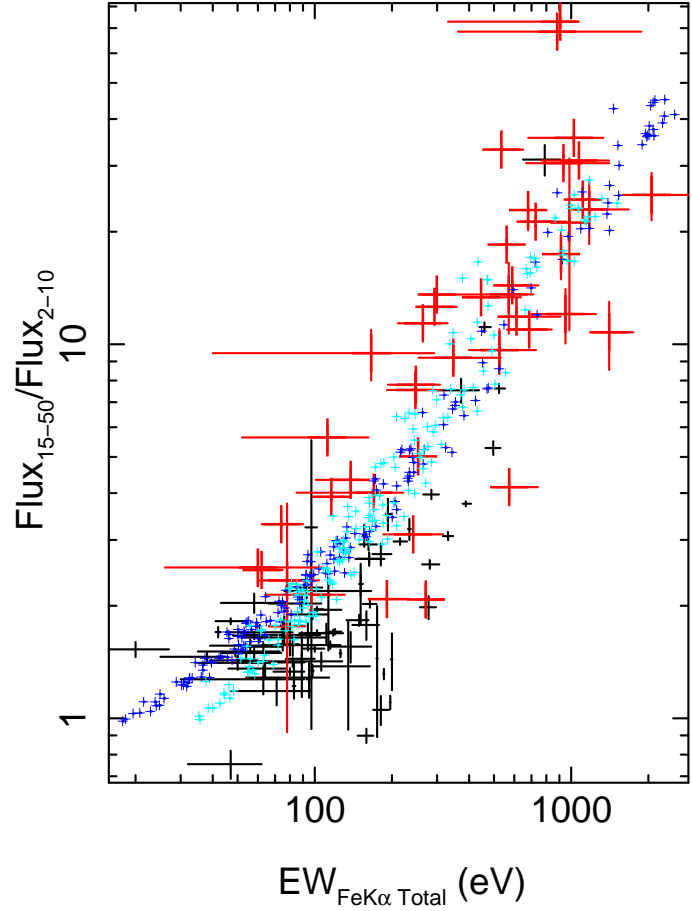}}}\,\,\,\,\,\,\,\,\,\,\,\,
\caption{ The hardness ratio plotted against the total Fe K$\alpha$ equivalent width, including both type 1 AGN (black markers) and type 2 AGN (red markers). Model overlaid are MCRT calculations assuming 1000 clouds in the spherical distribution, an inner radius of 10 units, an outer radius of 20 units, $\Gamma$=2.0, sin($\theta$) cloud distribution, and N$_H$ = 9 x 10$^{23}$ (cyan crosses) and 2 x 10$^{24}$ cm$^{-2}$ (blue crosses).}
\label{fig:EW}
\end{figure}

\begin{figure}
\epsscale{.5}
\plotone{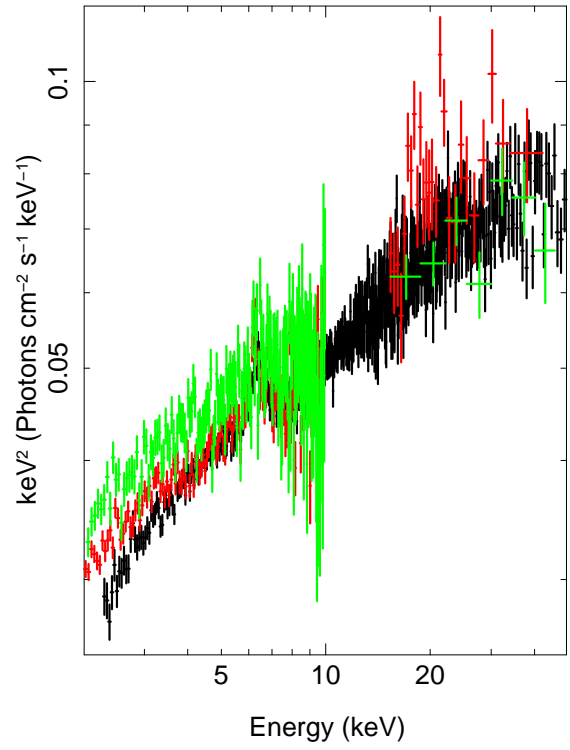}
\caption{ Spectra of the August 12, 2012 {\it NuSTAR} observation (black), the August 1, 2007 {\it Suzaku} observation (red) and the July 21, 1998 {\it BeppoSAX} observation (green) of IC4329A.}
\label{fig:Beppo_Suzaku}
\end{figure}

\begin{figure}
\center
\includegraphics[width=11cm]{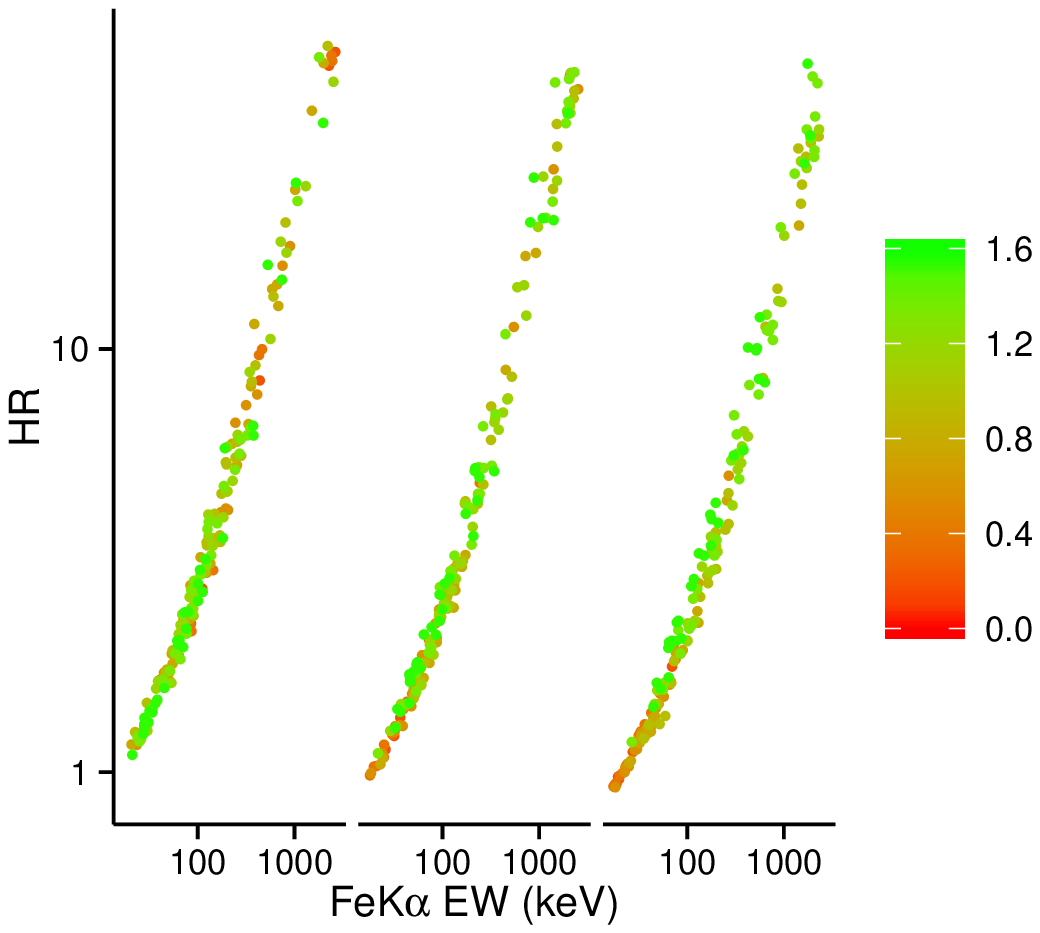}
\includegraphics[width=11cm]{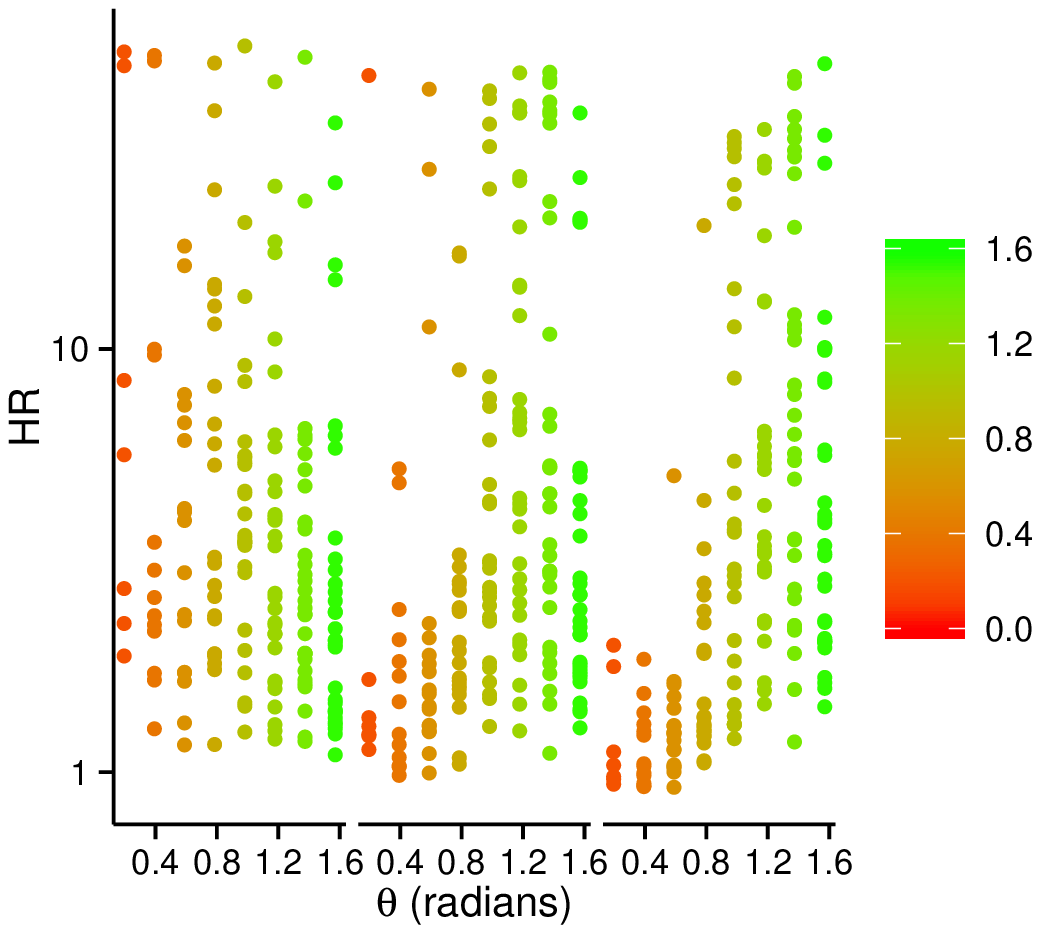}
\caption{ : MCRT simulated hardness ratio (HR) plotted against Fe K$\alpha$ equivalent width (EW) (top) and the MCRT simulated hardness ratio plotted against polar viewing angle ($\theta$) (bottom) with varying degrees of anisotropy. The spatial density of cloud centers is proportional to constant (left), sin$\theta$ (middle), and sin$^2$$\theta$ (right), and the colors indicate the polar angle of a sightline in radians.}
\label{fig:sinpower}
\end{figure}

\begin{figure}
\centering
\plotone{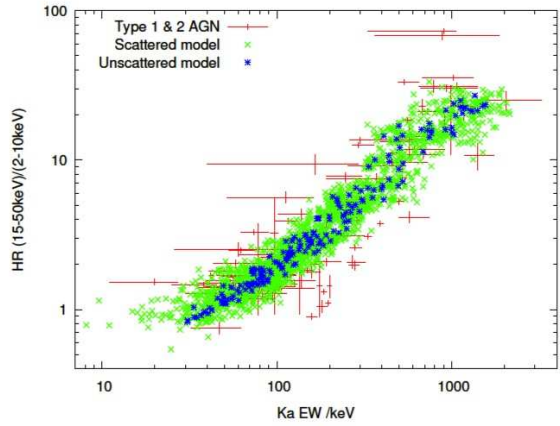}
\caption{ A subset of the MCRT model predictions with (green) and without (blue) error-scatter overlay  the equivalent width distribution of the type 1 and type 2 AGN (red). The models assumed N$_H$ = $10^{24}$ cm$^{-2}$, $\Gamma$ = 2.1 and outer radius=20 cloud radii.}
\label{fig:errors}
\end{figure}

\begin{figure}
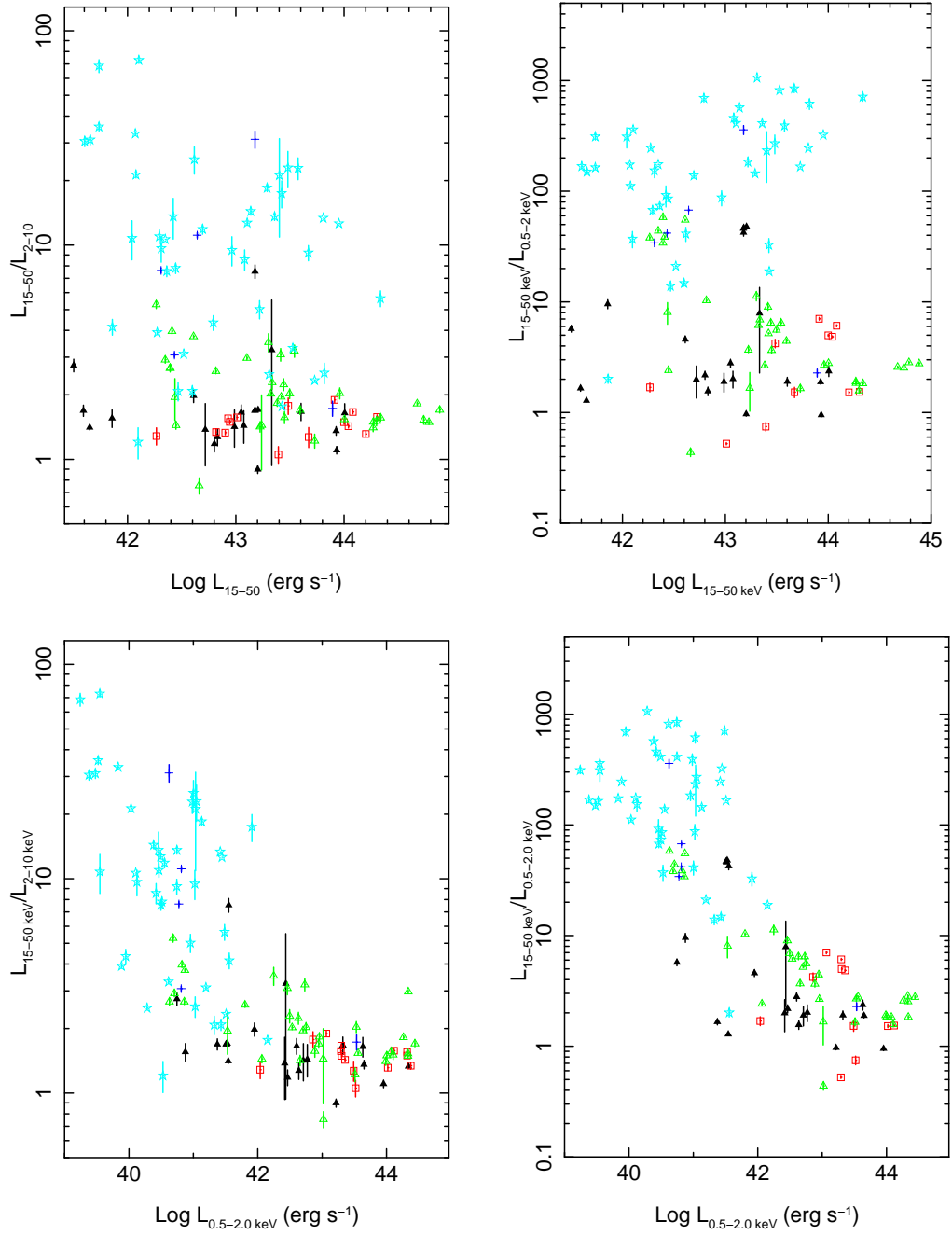

\epsscale{1}
\centering
\includegraphics[width=7cm,height=9cm]{lumfig1a.eps}
\includegraphics[width=7cm,height=9cm]{lumfig1b.eps}
\includegraphics[width=7cm,height=9cm]{lumfig2a.eps}
\includegraphics[width=7cm,height=9cm]{lumfig2b.eps}
\caption{Top: the 15-50/2-10 keV luminosity ratio (left) and 15-50/0.5-2 keV luminosity ratio (right) plotted against the 15-50 keV luminosity. 
Bottom: the 15-50/2-10 keV luminosity ratio (left) and 15-50/0.5-2 keV luminosity ratio (right) plotted against the 0.5-2 keV luminosity. With 
type 1 (filled black triangles), type 1.2 (red squares), type 1.5 (open green triangles), type 1.8-1.9 (dark blue crosses) and type 2 AGN (aqua stars).} 
\label{fig:lum_lum}
\end{figure}

\begin{deluxetable}{lccccclcc}
\tabletypesize{\scriptsize}
\tablecaption{X-ray Observation Type 2 Source List} 
\tablewidth{0pt}
\tablehead{
\colhead{Object} &
\colhead{ObsID}&
\colhead{RA (h m s)\tablenotemark{1}}&
\colhead{Dec ({$^\circ$ \arcmin\, \arcsec)}\tablenotemark{1}}&
\colhead{z\tablenotemark{1}}&
\colhead{N$_H$(Gal)\tablenotemark{2}}&
\colhead{Type\tablenotemark{1}}&
\colhead{Hardness Ratio\tablenotemark{3}}&
\colhead{Total EW (eV)\tablenotemark{4}}\\
}
\startdata
\multicolumn{9}{c}{}\\
NGC 4138\tablenotemark{5}&704047010&12 09 29.8&+ 43 41 07&0.003&0.014&1.9&2.19 $\pm{ 0.40 }$ &151 $^{+	15	}	_{-	75	}$\\
2MASXJ02485937+2630391	&	704013010	&	02 48 59.3	&+	26 30 39 	&	0.058	&	0.104	&	2	&	5.64	$\pm{	0.48	}$&	112	$^{+	50	}	_{-	60	}$	\\
2MASXJ04440903+2813003	&	703021010	&	04 44 09.0	&+	28 13 01	&	0.011	&	0.196	&	2	&	4.34	$\pm{	0.34	}$&	138	$^{+	14	}	_{-	37	}$	\\
2MASXJ12005792+0648226	&	703009010	&	12 00 57.9	&+	06 48 23	&	0.036	&	0.014	&	2	&	2.53	$\pm{	0.28	}$&	60	$^{+	48	}	_{-	34	}$	\\
2MASXJ20183871+4041003	&	506018010	&	20 18 38.7	&+	40 41 00	&	0.014	&	1.200	&	2	&	8.56	$\pm{	0.93	}$&	65	$^{+	71	}	_{-	44	}$	\\
Ark347	&	705002010	&	12 04 29.7	&+	20 18 58	&	0.022	&	0.024	&	2	&	9.47	$\pm{	1.47	}$&	166	$^{+	126	}	_{-	126	}$	\\
ESO005-G004	&	701018010	&	06 05 41.6	&--	86 37 55	&	0.006	&	0.111	&	2	&	10.77	$\pm{	2.23	}$&	1407	$^{+	341	}	_{-	221	}$	\\
ESO103-G035	&	703031010	&	18 38 20.3 	&--	65 25 39 	&	0.013	&	0.076	&	2	&	3.31	$\pm{	0.07	}$&	74	$^{+	16	}	_{-	12	}$	\\
ESO137-G034	&	403075010	&	16 35 14.1 	&-- 	58 04 48 	&	0.009	&	0.247	&	2	&	1.20	$\pm{	0.20	}$&	950	$^{+	300	}	_{-	250	}$	\\
ESO 297-018&701015010&01 83 37.1&-- 40 00 41&0.025&0.020&2&11.38 $\pm{ 0.30 }$ &264 $^{+	66	}	_{-	53	}$\\
IGRJ12391-1612	&	703007010	&	12 39 06.3	&-- 	16 10 47 	&	0.037	&	0.037	&	2	&	2.34	$\pm{	0.01	}$&	78	$^{+	26	}	_{-	17	}$	\\
MCG+4-48-002	&	702081010	&	20 20 35.0	&+	25 44 00	&	0.014	&	0.260	&	2	&	12.69	$\pm{	0.34	}$&	648	$^{+	122	}	_{-	73	}$	\\
MCG -01-05-047&704043010&01 52 49.0&-- 03 26 49&0.017&0.025&2&4.01 $\pm{ 0.10 }$ &170 $^{+ 51	}	_{- 85	}$	 \\ 
MCG -02-08-014&704045010&02 52 23.4&-- 08 30 37& 0.017&0.046&2&2.14 $\pm{ 0.02 }$ &97  $^{+ 34	}	_{- 33	}$	\\
MCG-5-23-16	&	700002010	&	09 47 40.1	&--	30 56 55	&	0.008	&	0.080	&	2	&	1.76	$\pm{	0.03	}$&	78	$^{+	6	}	_{-	7	}$	\\
Mrk18	&	705001010	&	09 01 58.4	&+	60 09 06 	&	0.011	&	0.044	&	2	&	10.64	$\pm{	0.01	}$&	253	$^{+	129	}	_{-	100	}$	\\
NGC1068	&	701039010	&	20 42 40.7 	&--	00 00 48	&	0.004	&	0.035	&	2	&	4.15	$\pm{	0.35	}$&	573	$^{+	169	}	_{-	71	}$	\\
NGC1142	&	701013010	&	02 55 12.2 	&--	00 11 01	&	0.029	&	0.064	&	2	&	12.59	$\pm{	0.00	}$&	293	$^{+	65	}	_{-	22	}$	\\
NGC1142	&	702079010	&	02 55 12.2 	&--	00 11 01	&	0.029	&	0.064	&	2	&	13.36	$\pm{	0.00	}$&	445	$^{+	193	}	_{-	54	}$	\\
NGC2992	&	700005010	&	09 45 42.0	&--	14 19 35 	&	0.008	&	0.053	&	2	&	2.08	$\pm{	0.20	}$&	271	$^{+	49	}	_{-	19	}$	\\
NGC2992	&	700005020	&	09 45 42.0	&--	14 19 35 	&	0.008	&	0.053	&	2	&	2.08	$\pm{	0.13	}$&	191	$^{+	43	}	_{-	26	}$	\\
NGC2992	&	700005030	&	09 45 42.0	&--	14 19 35 	&	0.008	&	0.053	&	2	&	3.10	$\pm{	0.05	}$&	242	$^{+	74	}	_{-	57	}$	\\
NGC3081	&	703013010	&	09 59 29.5	&--	22 49 35	&	0.008	&	0.046	&	2	&	11.86	$\pm{	0.62	}$&	612	$^{+	296	}	_{-	75	}$	\\
NGC3281	&	703033010	&	10 31 52.1	&--	34 51 13	&	0.011	&	0.064	&	2	&	14.37	$\pm{	0.58	}$&	589	$^{+	156	}	_{-	66	}$	\\
NGC3393	&	702004010	&	10 48 23.4	&--	25 09 43	&	0.013	&	0.061	&	2	&	25.09	$\pm{	3.65	}$&	2059	$^{+	1227	}	_{-	477	}$	\\
NGC4388	&	800017010	&	12 25 46.7	&+	12 39 44	&	0.008	&	0.026	&	2	&	5.01	$\pm{	0.49	}$&	253	$^{+	11	}	_{-	10	}$	\\
NGC4507	&	702048010	&	12 35 36.6	&--	39 54 33	&	0.012	&	0.072	&	2	&	18.48	$\pm{	0.72	}$&	560	$^{+	28	}	_{-	30	}$	\\
NGC454	&	704009010	&	01 14 22.5	&--	55 23 55	&	0.012	&	0.027	&	2	&	9.65	$\pm{	1.33	}$&	525	$^{+	205	}	_{-	126	}$	\\
NGC4945	&	100008010	&	13 05 27.5	&--	49 28 06	&	0.002	&	0.157	&	2	&	21.27	$\pm{	0.70	}$&	727	$^{+	228	}	_{-	82	}$	\\
NGC4945	&	100008030	&	13 05 27.5	&--	49 28 06	&	0.002	&	0.157	&	2	&	33.14	$\pm{	0.62	}$&	534	$^{+	117	}	_{-	41	}$	\\
NGC4945	&	705047010	&	13 05 27.5	&--	49 28 06	&	0.002	&	0.157	&	2	&	68.55	$\pm{	4.76	}$&	882	$^{+	996	}	_{-	520	}$	\\
NGC4945	&	705047020	&	13 05 27.5	&--	49 28 06	&	0.002	&	0.157	&	2	&	35.64	$\pm{	1.55	}$&	1024	$^{+	313	}	_{-	344	}$	\\
NGC4945	&	705047030	&	13 05 27.5	&--	49 28 06	&	0.002	&	0.157	&	2	&	72.87	$\pm{	2.47	}$&	907	$^{+	153	}	_{-	576	}$	\\
NGC4945	&	705047040	&	13 05 27.5	&--	49 28 06	&	0.002	&	0.157	&	2	&	31.00	$\pm{	1.65	}$&	1072	$^{+	334	}	_{-	306	}$	\\
NGC4945	&	705047050	&	13 05 27.5	&--	49 28 06	&	0.002	&	0.157	&	2	&	30.50	$\pm{	1.68	}$&	933	$^{+	475	}	_{-	266	}$	\\
NGC 5728&701079010&14 42 23.9& -- 17 15 11& 0.009&0.078&2&24.37 $\pm{ 0.40 }$&1110 $^{+	111	}	_{-	111	}$	\\
NGC6300	&	702049010	&	17 16 59.5	&--	62 49 14	&	0.004	&	0.099	&	2	&	3.91	$\pm{	0.09	}$&	116	$^{+	14	}	_{-	18	}$	\\
NGC6552	&	504070010	&	18 00 07.3 	&+	66 36 54	&	0.026	&	0.042	&	2	&	22.94	$\pm{	4.43	}$&	1176	$^{+	502	}	_{-	141	}$	\\
NGC6552	&	504072010	&	18 00 07.3 	&+	66 36 54	&	0.026	&	0.042	&	2	&	22.85	$\pm{	2.67	}$&	679	$^{+	0	}	_{-	0	}$	\\
NGC6552	&	504074010	&	18 00 07.3 	&+	66 36 54	&	0.026	&	0.042	&	2	&	17.43	$\pm{	2.52	}$&	913	$^{+	0	}	_{-	0	}$	\\
NGC6552	&	504076010	&	18 00 07.3 	&+	66 36 54	&	0.026	&	0.042	&	2	&	21.15	$\pm{	10.25	}$&	984	$^{+	0	}	_{-	0	}$	\\
NGC7172	&	703030010	&	22 02 01.9	&--	31 52 11	&	0.009	&	0.017	&	2	&	2.49	$\pm{	0.04	}$&	62	$^{+	13	}	_{-	9	}$	\\
NGC7582	&	702052010	&	23 18 23.5	&--	42 22 14	&	0.005	&	0.019	&	2	&	7.80	$\pm{	0.43	}$&	248	$^{+	60	}	_{-	55	}$	\\
NGC7582	&	702052020	&	23 18 23.5	&--	42 22 14	&	0.005	&	0.019	&	2	&	7.55	$\pm{	0.43	}$&	247	$^{+	65	}	_{-	56	}$	\\
NGC7582	&	702052030	&	23 18 23.5	&--	42 22 14	&	0.005	&	0.019	&	2	&	13.59	$\pm{	2.92	}$&	571	$^{+	142	}	_{-	102	}$	\\
NGC7582	&	702052040	&	23 18 23.5	&--	42 22 14	&	0.005	&	0.019	&	2	&	10.95	$\pm{	0.83	}$&	686	$^{+	154	}	_{-	124	}$	\\
NGC788	&	703032010	&	02 01 06.4	&--	06 48 56 	&	0.014	&	0.022	&	2	&	13.58	$\pm{	0.55	}$&	299	$^{+	169	}	_{-	38	}$	\\
SWIFTJ0138.6-4001	&	701015010	&	01 38 37.1	&--	40 00 41 	&	0.025	&	0.020	&	2	&	9.20	$\pm{	0.75	}$&	347	$^{+	184	}	_{-	93	}$	\\
\enddata 
\tablenotetext{1}{ \citet{veron06}}
\tablenotetext{2}{The Galactic column density in units of 10$^{22}$ cm$^{-2}$ obtained from the weighted average $N_H$ in the Dickey \& Lockman HI in the Galaxy survey \citep{Dickey:1990uq}}
\tablenotetext{3}{ Flux$_{15-50\,\textnormal{keV}}$/Flux$_{2-10\,\textnormal{keV}}$ with errors calculated from the net count rate errors of the XIS and PIN data}
\tablenotetext{4}{ Total equivalent width of the Fe K$\alpha$ emission line with the errors to 90\% confidence}
\tablenotetext{5}{ See Section \ref{sec:sources} for details}
\label{tab:table}
\end{deluxetable}
\end{document}